\documentclass[
    pra,  
    twocolumn, 
    floatfix, 
    amsmath, 
    amssymb, 
    superscriptaddress
]{revtex4-1}
\usepackage{graphicx}% Include figure files
\usepackage{bm}% bold math
\begin{document}

%\preprint{APS/123-QED}

\title{Many-particle Entanglement in Multiple Quantum NMR Spectroscopy} 

\author{S.I. Doronin}
\author{E.B Fel'dman}
\affiliation{ Institute of Problems of Chemical Physics of Russian Academy of Sciences, \\ Chernogolovka, Moscow Region, Russia 142432
}
\author{I.D. Lazarev}
\affiliation{ Institute of Problems of Chemical Physics of Russian Academy of Sciences, \\ Chernogolovka, Moscow Region, Russia 142432
}

\affiliation{ Faculty of Fundamental Physical-Chemical Engineering, Lomonosov Moscow State University, GSP-1, Moscow, Russia 119991
}

%\date{\today}

\begin{abstract}
We use multiple quantum (MQ) NMR dynamics of a gas of spin-carrying molecules in nanocavities at high and low temperatures for an investigation of many-particle entanglement. 
A distribution of MQ NMR intensities is obtained at high and low temperatures in a system  of 201 spins $\frac 1 2$. 
The second moment of the distribution, which provides a lower bound on the quantum Fisher information, sheds light on the many-particle entanglement in the system. 
The dependence of the many-particle entanglement on the temperature is investigated.  
Almost all spins are entangled at low temperatures.

%\begin{description}
%\end{description}
\end{abstract}

\maketitle

%\tableofcontents
\section{Introduction}
Multiple quantum (MQ) NMR spectroscopy \cite{mq_nmr_experiment} was introduced for the investigation of nuclear spin distributions in various materials (liquid crystals \cite{spin_distribution_in_liquid_system}, simple organic systems \cite{mq_nmr_experiment}, amorphous hydrogenated silicon \cite{spin_distribution_in_silicon}, etc.). 
It also turned out to be useful for probing the decoherence rate in highly correlated spin clusters \cite{decoherence_register,decoherence_ca_f2}. 
The scaling of the decoherence rate with the number of the correlated spins has also been demonstrated \cite{decoherence_register,lab:decoherence_2018}. 
Essentially, the MQ NMR dynamics is a suitable method to quantify the development of MQ coherences starting from the $z$-polarisation and ending with a collective state of all spins. 
The method allows us to describe the spreading of correlations \cite{mq_nmr_experiment,spin_distribution_in_liquid_system,decoherence_under_dq,nmr_dyn} and offers a signature of localisation effects \cite{loc_deloc_nmr_dyn,loc_in_chain}. The spreading rate can be described through out-of-time ordered correlations (OTOCs) which are connected with the distribution of MQ NMR coherences. 
\par
Attempts to quantify entanglement are motivated by the
desire to understand and quantify resources responsible for
advantages  of quantum computing over classical computing.
Pair entanglement is the most familiar while many-particle entanglement is its most general extension.
The existing applications of  MQ NMR to quantum information make it important to understand many-particle entanglement in the MQ NMR context.  
The starting point of such investigation must be the simplest model with non-trivial behavior. This is the motivation of the present work.
\par
Connections between MQ coherences and entanglement have been established only for spin pairs \cite{lab:entanglement_dyn_2003,entanglement_dyn,nuclear_polarization_and_entanglement}.  The same is true for the MQ coherences as a witness of entanglement \cite{sep_of_mixed_states}. The MQ NMR coherence of the second order was used for the construction of an entanglement witness for a two-spin system with the dipole-dipole interactions (DDIs) \cite{lab:entanglement_witness_nmr_2008,lab:mq_mnr_qinfo_2012}. At the same time, MQ NMR dynamics allows us to clarify deeper connections between MQ coherences and entanglement. Those connections are closely related to the spread of MQ correlations inside a many-spin system in the evolution process. As a result, it is possible to extract information about many-qubit entanglement and entanglement witnesses from the second moment of the intensity spectrum of the MQ NMR coherences \cite{otoc_to_enanglement_via_mqcoh}. It is also important that there is a relationship between the second moment of MQ NMR coherences and the quantum Fisher information \cite{qmetrology_for_qinfo,qmetrology_nonclassiscal_state}. In particular, it was shown that the second moment of the MQ NMR spectrum provides a lower bound on the quantum Fisher information \cite{otoc_to_enanglement_via_mqcoh}.
\par
In order to investigate many-spin entanglement it is necessary to work out a model of interacting spins, in which many-spin dynamics can be studied at low temperatures. 
It is also important that the model contains a sufficiently large number of spins and is applicable at arbitrary temperatures.
Only then it is possible to investigate many-spin entanglement and its dependence on the temperature.
\par
One would think that MQ NMR in one-dimensional systems is most suitable for the investigation of many-spin entanglement because a consistent quantum-mechanical theory of MQ NMR dynamics has been developed only for one-dimensional systems \cite{lab:nmr_dyn_1996,lab:mq_nmr_in_chain_1997,lab:mq_dyn_of_chain_in_solid_2000}. However, this is not the case. The point is that the exact solutions for MQ NMR dynamics of one-dimensional systems demonstrate \cite{lab:nmr_dyn_1996,lab:mq_nmr_in_chain_1997,lab:mq_dyn_of_chain_in_solid_2000} that, starting from a thermodynamic equilibrium state, only zero and double quantum coherences are produced in the approximation of the nearest neighbour interactions. As a result, the second moment (dispersion) of the MQ NMR spectrum is small and many-qubit entanglement does not appear.
\par
For the investigation of many-qubit entanglement, we build on the model \cite{nanopore_model} of a non-spherical nanopore filled with a gas of spin-carrying atoms (for example, xenon) or molecules in a strong external magnetic field. 
It is well known that the dipole-dipole interactions (DDIs) of spin-carrying atoms (molecules) in such nanopores do not average out to zero due to molecular diffusion \cite{nanopore_model,lab:depolar_in_nanocavities_2004}. 
It is very significant that the residual averaged DDIs are determined by only one coupling constant, which is the same for all pairs of interacting spins \cite{nanopore_model,lab:depolar_in_nanocavities_2004}. 
This means that essentially we have a system of equivalent spins and its MQ NMR dynamics can be investigated exactly \cite{lab:mq_nmr_dyn_in_nanopores_2009}.
For this model OTOCs allow the investigation of many-particle entanglement, and the extraction of information about the number of the entangled spins during the system evolution. 
The temperature dependence of the number of the entangled spins can be also investigated.
We discover that the system exhibits $k$-spin entanglement with $k$ growing as the temperature decreases. 
Almost all spins are entangled at low temperatures despite the absence of  entanglement in the initial state. We expect this behavior to be generic for MQ NMR.
\par
The existing theoretical approach \cite{lab:mq_nmr_dyn_in_nanopores_2009} to the MQ NMR dynamics of a system of equivalent spins is valid only in the high-temperature region. In order to investigate the many-qubit entanglement, we develop a theory of the MQ NMR dynamics of equivalent spins at low temperatures. 
We perform all calculations for a system of 201 spins. 
Such an investigation of many-spin entanglement is performed for the first time.
In principle, analogous calculations can be performed for  systems with several thousand  spins. 
\par
The present paper investigates the connection of the second moment of the MQ NMR spectrum of spin-carrying atoms (molecules) in a nanopore with many-spin entanglement in the system. The paper is organised as follows. In Sec. \ref{sec:mq_dyn} the theory of MQ NMR dynamics at low temperatures in the system of equivalent spins coupled by the DDIs is developed. An analytical solution for the MQ NMR dynamics of a three-spin system is obtained in Sec. \ref{sec:exact_sol}. The second moment (dispersion) of the MQ NMR spectrum of the system of equivalent spins in a nanopore at arbitrary temperatures is obtained in Sec. \ref{sec:second_moment}. The investigation of the dependence of  many-spin entanglement on the temperature is given in Sec. \ref{sec:entanglement}. We briefly summarise our results in the concluding Sec. \ref{sec:conslusions}.

\section{MQ NMR dynamics of spins-1/2 in a nanopore at low temperatures}
\label{sec:mq_dyn}

\begin{figure}
    \centering
    \includegraphics[width=\linewidth]{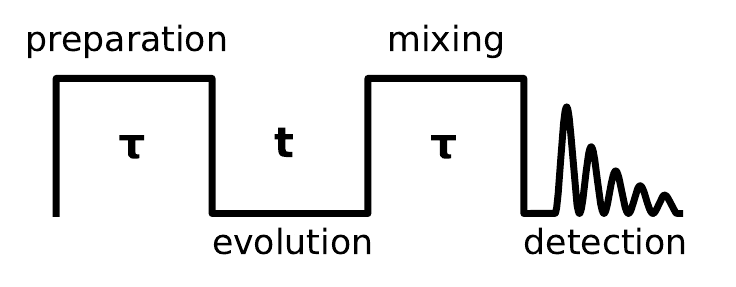}
    \caption{The basic scheme of the MQ NMR experiment.}
    \label{fig:experiment}
\end{figure}

The standard MQ NMR experiment consists of four distinct periods of time (Fig.~\ref{fig:experiment}): preparation ($\tau$), evolution ($t_1$), mixing ($\tau$) and detection ($t_2$) \cite{mq_nmr_experiment}.
MQ coherences are created by a multi-pulse sequence irradiating the system on the preparation period \cite{mq_nmr_experiment}. Since the correlation time of the molecular diffusion of spin-carrying atoms (molecules) in nanopores is much shorter than both the dipolar time $t \approx \omega^{-1}_{\mathrm{loc}}$ ( $\omega^{-1}_{\mathrm{loc}}$  is the dipolar local field in the frequency units \cite{Goldman}) and  the period of the multi-pulse sequence on the preparation period of the MQ NMR experiment \cite{mq_nmr_experiment}, one can assume that spin dynamics is governed by the averaged dipolar coupling constant, $D$, which is the same for all spin pairs. Then the averaged non-secular two-spin/two-quantum Hamiltonian, $H_\mathrm{MQ}$, describing MQ dynamics on the preparation period, can be written in the rotating reference frame \cite{Goldman} as \cite{lab:mq_nmr_dyn_in_nanopores_2009}

\begin{equation}
    \label{eq:ham_mq}
    H_\mathrm{MQ}  = - \frac D 4 \left\{(I^+)^2 + (I^-)^2\right\},
\end{equation}
where  $I^{\pm} = \sum\limits_{j=1}^N I^{\pm}_j$, $N$ is the number of the spins in the nanopore, and $I^{\pm}_j$ are the raising/lowering operators of spin j.

In order to investigate the MQ NMR dynamics of the system one should find the density matrix $\rho(\tau)$ on the preparation period of the MQ NMR experiment \cite{mq_nmr_experiment} by solving the Liouville evolution equation \cite{lab:mq_mnr_qinfo_2012}

\begin{equation}
    \label{eq:liouvile}
    i \dfrac{d\rho(\tau)}{d \tau} = 
    \left[H_\mathrm{MQ}, \rho(\tau)\right]
\end{equation}
with the initial thermodynamic equilibrium density matrix

\begin{equation}
    \label{eq:rho_eq}
    \rho(0) = \rho_{\mathrm{eq}} = 
    \dfrac{e^{\frac{\hbar\omega_{0}}{kT} I_z}}{Z},
\end{equation}
where $Z =\mathrm{Tr}\left\{e^{\frac{\hbar\omega_{0}}{kT} I_z}\right\}$ is the partition function, $\hbar$ and $k$ are the Plank and Boltzmann constants, $\omega_0$ is the Larmor frequency, $T$ is the temperature, and $I_z$ is the operator of the projection of the total spin angular momentum on the $z$-axis, which is directed along the strong external magnetic field. In the high-temperature approximation \cite{Goldman}, when $b = \frac{\hbar\omega_{0}}{kT} \ll 1$, we can rewrite Eq.   (\ref{eq:rho_eq}) as 

\begin{equation}
    \label{eq:rho_ht}
    \rho(0) = \rho_{\mathrm{eq}} \approx
    \dfrac{1}{2^N} (1 + bI_z).
\end{equation}

Following the preparation, evolution, and mixing periods of the MQ NMR experiment \cite{mq_nmr_experiment}, the resulting signal $G(\tau, \phi)$ stored as population information is \cite{lab:low_temp_dyn_1997}

\begin{multline}
    \label{eq:signal}
     G(\tau, \phi) = \\ 
     \mathrm{Tr}\left\{
         e^{iH_\mathrm{MQ}\tau} e^{i\phi I_z} e^{-iH_\mathrm{MQ}\tau}
         \rho_\mathrm{eq}
         e^{iH_\mathrm{MQ}\tau} e^{-i\phi I_z} e^{-iH_\mathrm{MQ}\tau}
         I_z \right\} = \\
    \mathrm{Tr}\left\{e^{i\phi I_z} \rho_\mathrm{LT} (\tau)
              e^{-i\phi I_z} \rho_\mathrm{HT} (\tau) \right\},
\end{multline}
where
\begin{align}
    \label{eq:eval_rho}
    \rho_\mathrm{LT} (\tau) & = e^{-iH_\mathrm{MQ}\tau} \rho_\mathrm{eq} e^{iH_\mathrm{MQ}\tau},
    \notag\\
    \rho_\mathrm{HT} (\tau) & =  e^{-iH_\mathrm{MQ}\tau} I_z e^{iH_\mathrm{MQ}\tau}.
\end{align}

It proves convenient to expand the spin density matrices, $\rho_\mathrm{LT}(\tau)$ and $\rho_\mathrm{HT}(\tau)$, in series as

\begin{equation}
    \label{eq:rho_series}
    \rho_\mathrm{LT} (\tau) = \sum_n \rho_{LT, n}(\tau); \quad
    \rho_\mathrm{HT} (\tau) = \sum_n \rho_{HT, n}(\tau),
\end{equation}
where $\rho_{LT, n} (\tau)$ and $\rho_{HT, n} (\tau)$ are the contributions to $\rho_\mathrm{LT} (\tau)$ and $\rho_\mathrm{HT} (\tau)$ from the MQ coherence of the $n$th order. Then the resulting signal $G(\tau, \phi)$ of the MQ NMR \cite{mq_nmr_experiment} can be rewritten as 

\begin{equation}
    \label{eq:signal_series}
    G(\tau, \phi) = \sum\limits_n 
    e^{in\phi}\mathrm{Tr} \left\{
    \rho_{LT, n}(\tau) \rho_{HT, -n} (\tau)
    \right\},
\end{equation}
where we took into account that
\begin{equation}
    \left[I_z, \rho_{LT, n} (\tau) \right] = n  \rho_{LT, n} (\tau);
    \quad 
    \left[I_z, \rho_{HT, n} (\tau) \right] = n  \rho_{HT, n} (\tau).
\end{equation}
The normalised intensities of the MQ NMR coherences can be expressed as follows 

\begin{equation}
    \label{eq:coherence}
    J_n(\tau) = 
    \dfrac{
       \mathrm{Tr} \left\{
        \rho_{LT, n}(\tau) \rho_{HT, -n} (\tau)
        \right\}
    }{Tr \left\{\rho_{\mathrm{eq}}I_z\right\}}.
\end{equation}
A simple calculation using Eq.   (\ref{eq:rho_eq}) yields \cite{lab:low_temp_dyn_1997}

\begin{equation}
   \mathrm{Tr} \left\{\rho_{\mathrm{eq}}I_z\right\} = 
    \frac N 2 \tanh \frac b 2.
\end{equation}
The normalised intensity $J_0(0)$ of the MQ NMR coherence of the zeroth order at $\tau=0$ equals 1 and all the other intensities are zero. Using Eqs.   (\ref{eq:rho_eq}, \ref{eq:eval_rho}) one can find that

\begin{equation}
    \rho_\mathrm{LT}(\tau) = \dfrac 1 Z 
    \exp\left(be^{-iH_\mathrm{MQ}\tau} I_z e^{iH_\mathrm{MQ}\tau}\right) =
    \dfrac 1 Z e^{b\rho_\mathrm{HT}(\tau)}.
\end{equation}
Further,
\begin{multline}
    \label{eq:sum_of_coherence}
    \sum\limits_n J_n(\tau) = 
    \dfrac{\sum\limits_{n, m}\mathrm{Tr}\left\{
        \rho_{LT, n}(\tau)\rho_{HT, m}(\tau)
    \right\}}
    {\mathrm{Tr}\left\{\rho_\mathrm{eq} I_z\right\}} = \\
    \dfrac{\mathrm{Tr}\left\{
        \rho_\mathrm{LT}(\tau)\rho_\mathrm{HT}(\tau)
    \right\}}
    {\mathrm{Tr}\left\{\rho_\mathrm{eq} I_z\right\}} = 
    \dfrac{\mathrm{Tr}\left\{
        \rho_\mathrm{HT}(\tau)e^{b\rho_\mathrm{HT}(\tau)}
    \right\}}
    {Z\mathrm{Tr}\left\{\rho_\mathrm{eq} I_z\right\}} = \\
    \dfrac{\frac{d}{db} \ln \mathrm{Tr}\left\{
        e^{b I_z}
    \right\}}
    {\mathrm{Tr}\left\{\rho_\mathrm{eq} I_z\right\}} = 
    \dfrac{\frac 1 2 N \tanh \left( \frac b 2 \right)}
    {\frac 1 2 N \tanh \left( \frac b 2 \right)} = 1.
\end{multline}
Eq.   (\ref{eq:sum_of_coherence}) means that the sum of the MQ NMR coherences is conserved on the preparation period of the MQ NMR experiment \cite{mq_nmr_experiment}.

The Hamiltonian $H_\mathrm{MQ}$ of Eq.   (\ref{eq:ham_mq}) commutes with the square of the total spin angular momentum $\hat I^2$ and we will use the basis consisting of the common eigenstates of $(\hat I)^2$ and $I_z$ to study MQ NMR dynamics as done in Ref.\cite{lab:mq_nmr_dyn_in_nanopores_2009} at high temperatures. In this basis, the Hamiltonian $H_{MQ}$ consists of blocks $H_{MQ}^S$, corresponding to different values of the total spin angular momentum $S$ ($\hat I^2 = S(S+1), S = N/2, N/2-1, N/2-2, \dots, N/2 - [N/2]$, $[i]$ is the integer part of $i$).
Since both the Hamiltonian $H_{MQ}$  and the initial density matrix of Eq.   (\ref{eq:rho_eq}) exhibit block structure, one can conclude that the density matrices $\rho_\mathrm{LT}(\tau)$ and $\rho_\mathrm{HT}(\tau)$ consist of blocks $\rho^S_\mathrm{LT}(\tau)$ and $\rho^S_\mathrm{HT}(\tau)$ $(S=\frac N 2, \frac N 2 - 1, \dots, \frac N 2 - \left[\frac N 2\right])$ as well. We will denote as $\rho^S_{LT, n}(\tau)$ and $\rho^S_{HT, n}(\tau)$ the contributions to $\rho^S_\mathrm{LT}(\tau)$ and $\rho^S_\mathrm{HT}(\tau)$ from the MQ coherence of order $n$. Then the contribution $J_{n, S}(\tau)$ to the intensity of the $n$-th order MQ NMR coherence is determined as 

\begin{equation}
    \label{eq:coherence_k_s}
    J_{n, S}(\tau) = \dfrac{\mathrm{Tr}\left\{
        \rho_{LT, n}^S(\tau)\rho_{HT, -n}^S(\tau)
    \right\}}
    {\mathrm{Tr}\left\{\rho_{eq} I_z\right\}}.
\end{equation}
Thus, the problem is reduced to a set of analogous problems for each block $H_{MQ}^S$. The number of the states $n_N(S)$ of the total angular momentum $S$ in an $N$-spin system is \cite{Landau}

\begin{equation}
    \label{eq:coeff_n}
    n_N(S)  = \dfrac{ N! (2S+1)}
    {(\frac N 2 + S + 1)!(\frac N 2 - S)!}, 
    \quad
    0\leq S \leq \frac N 2, 
\end{equation}
which is also the multiplicity of the intensities $J_{n, S}(\tau)$. Then the observable intensities of the  MQ NMR coherences $J_n(\tau)\quad(-N\leq n \leq N)$ are 

\begin{equation}
    \label{eq:coherence_k}
    J_n(\tau) = \sum\limits_S n_N(S) J_{n, S}(\tau)
\end{equation}

The matrix representations of $(I^{\pm})^2$, which are necessary in order to find the Hamiltonian $H_{MQ}$ of (1) and to calculate $J_n(\tau)$, are given in \cite{lab:mq_nmr_dyn_in_nanopores_2009}. 

The dimension of the block $H_{MQ}^S$ is $2S+1$. One can verify \cite{lab:mq_nmr_dyn_in_nanopores_2009} that the total dimension of the Hamiltonian $H_{MQ}$ is 

\begin{equation}
    \sum\limits_N n_N(S)(2S+1) = 2^N.
\end{equation}

Since the Hamiltonian $H_{MQ}$ of Eq.   ~(\ref{eq:ham_mq}) commutes with the operator $e^{i\pi I_z}$, the $2^N\times2^N$ Hamiltonian matrix is reduced to two $2^{N-1}\times2^{N-1}$ submatrices \cite{lab:mq_nmr_dyn_in_nanopores_2009}. The same is valid for all blocks $H_{MQ}^S$ and $H_{MQ}^{-S}$. This reduction is valid both for even and odd $N$. For odd $N$, both submatrices give the same contribution to the MQ NMR coherences, and one should solve the problem using only one $2^{N-1}\times2^{N-1}$ submatrix and double the obtained intensities. In our calculations we take $N=201$.

\section{The exact solution for MQ NMR dynamics for a three-spin system in a nanopore at low temperatures}
\label{sec:exact_sol}

We consider a system of $N=3$ spins coupled through the Hamiltonian $H_{MQ}$ of Eq.   (\ref{eq:ham_mq}). The possible values $S$ of the total spin angular momentum are $\frac 3 2$ and $\frac 1 2$. One can find that the matrix representation of $H_{MQ}^{3/2}$  is 

\begin{equation}
    \label{eq:ham_3_2}
    H_{MQ}^{3/2} = 
    \begin{pmatrix}
        0 & 0 & -\frac{\sqrt{3} D}{2} & 0 \\
        0 & 0 & 0 & -\frac{\sqrt{3} D}{2} \\
        -\frac{\sqrt{3} D}{2} & 0 & 0 & 0 \\
        0 & -\frac{\sqrt{3} D}{2} & 0 & 0 
    \end{pmatrix}.
\end{equation}
The eigenvalues $\lambda_{3/2}^{(i)}(i=1, 2, 3, 4)$ of  $H_{MQ}^{3/2}$ are the following 
\begin{align}
    \label{eq:eigvals_3_2}
    \lambda_{3/2}^{(1)} &= -\frac{\sqrt{3} D}{2}, \quad
    \lambda_{3/2}^{(2)} = -\frac{\sqrt{3} D}{2}, \notag \\
    \lambda_{3/2}^{(3)} &= \frac{\sqrt{3} D}{2}, \quad
    \lambda_{3/2}^{(4)} = \frac{\sqrt{3} D}{2}.
\end{align}
The appropriate set of eigenvectors reads as follows:

\begin{align}
    \label{eq:eigvecs_3_2}
    u_{3/2}^{(1)} & =  \left(\frac{1}{\sqrt{2}}, 0, 
                   \frac{1}{\sqrt{2}}, 0\right) ,
    \notag \\
    u_{3/2}^{(2)} & =  \left(0, \frac{1}{\sqrt{2}}, 
                   0, \frac{1}{\sqrt{2}}\right) ,
    \notag \\
    u_{3/2}^{(3)} & =  \left(-\frac{1}{\sqrt{2}}, 0, 
                   \frac{1}{\sqrt{2}}, 0\right) ,
    \notag \\               
    u_{3/2}^{(4)} & =  \left(0, -\frac{1}{\sqrt{2}}, 
                   0, \frac{1}{\sqrt{2}}\right)  .              
\end{align}
The block $H^{1/2}_{MQ}$ is a scalar
\begin{equation}
    \label{eq:ham_1_2}
    H^{1/2}_{MQ} = 0.
\end{equation}
The solution of Eq.   (\ref{eq:liouvile}), $\rho_{\alpha}^{n/2} (\tau) \quad (n = 1, 3; \, \alpha = \mathrm{LT}, \mathrm{HT})$, where the Hamiltonian $H_{MQ}$ is replaced by the Hamiltonian $H_{MQ}^{n/2}$, is 

\begin{equation}
    \label{eq:liouvile_sol}
    \rho_{\alpha}^{n/2}(\tau) = 
    U_{n/2} e^{-i\Lambda^{n/2}\tau} U^{+}_{n/2}
    \rho^{n/2}_{\alpha}(0)
    U_{n/2} e^{i\Lambda^{n/2}\tau} U^{+}_{n/2},
\end{equation}
where $\Lambda^{n/2}$ is the diagonal matrix of the eigenvalues and $U_{n/2}$ is the matrix of the eigenvectors of the block $H_{MQ}^{n/2} \quad (n=1, 3)$, and the initial density matrices $\rho_\mathrm{LT}^{n/2}(0)(\alpha=LT)$ and $\rho_\mathrm{HT}^{n/2}(0)(\alpha=HT)$ are the following:

\begin{align}
    \label{eq:rho_LT_init}
    \rho_\mathrm{LT}^{3/2}(0) &= \dfrac 1 Z
    \begin{pmatrix}
        e^{\frac{3b}{2}} & 0 & 0 & 0 
        \\
        0 & e^{\frac{b}{2}} & 0 & 0 
        \\
        0 & 0 & e^{-\frac{b}{2}} & 0 
        \\
        0 & 0 & 0 & e^{-\frac{3b}{2}}
    \end{pmatrix}, 
    \notag \\
    \rho_\mathrm{LT}^{1/2}(0) &= \dfrac 1 Z
    \begin{pmatrix}
        e^{\frac{b}{2}} & 0 
        \\
        0 & e^{-\frac{b}{2}}
    \end{pmatrix},
\end{align}

\begin{align}
    \label{eq:rho_HT_init}
    \rho_\mathrm{HT}^{3/2}(0) & = 
    \begin{pmatrix}
        \frac{3}{2} & 0 & 0 & 0 
        \\
        0 & \frac{1}{2} & 0 & 0 
        \\
        0 & 0 & -\frac{1}{2} & 0 
        \\
        0 & 0 & 0 & -\frac{3}{2}
    \end{pmatrix},
    \notag \\
    \rho_\mathrm{HT}^{1/2}(0) &= 
    \begin{pmatrix}
        \frac{1}{2} & 0 
        \\
        0 & -\frac{1}{2}
    \end{pmatrix}.
\end{align}
After a calculation using Eqs.  (\ref{eq:eigvals_3_2}), (\ref{eq:eigvecs_3_2}), (\ref{eq:liouvile_sol}), (\ref{eq:rho_LT_init}), and (\ref{eq:rho_HT_init}) with $n = 3$, one obtains 
\begin{equation}
    \label{eq:rho_HT_eval_3_2}
    \rho_\mathrm{HT}^{3/2}(\tau) =
    \begin{pmatrix}
            \xi + \frac 1 2  
        & 
            0 
        & 
            -i\eta 
        & 
            0 
        \\
            0 
        & 
            \xi - \frac 1 2 
        & 
            0 
        & 
            -i\eta
        \\
            i\eta 
        &
            0 
        & 
            -\xi + \frac 1 2  
        & 
            0 
        \\
            0 
        & 
            i\eta 
        & 
            0 
        & 
            -\xi - \frac 1 2 
    \end{pmatrix},
\end{equation}
where 
\begin{equation}
    \xi = \cos(\sqrt{3}D\tau), \quad 
    \eta = \sin(\sqrt{3}D\tau), 
\end{equation}
and
\begin{multline}
    \label{eq:rho_LT_eval_3_2}
    \rho_\mathrm{LT}^{3/2}(\tau) = \frac{1}{Z} \\
    \begin{pmatrix}
            ue^{-\frac{b}{2}} + ve^{\frac{3b}{2}}
        & 
            0 
        & 
            -ie^{\frac{b}{2}}w
        & 
            0 
        \\
            0 
        & 
            ue^{-\frac{3b}{2}} + ve^{\frac{b}{2}} 
        & 
            0 
        & 
            -ie^{-\frac{b}{2}}w 
        \\
            ie^{\frac{b}{2}}w
        &
            0 
        & 
            ue^{\frac{3b}{2}} + ve^{-\frac{b}{2}}
        & 
            0 
        \\
            0 
        & 
            ie^{-\frac{b}{2}}w
        & 
            0 
        & 
            ue^{\frac{b}{2}} + ve^{-\frac{3b}{2}}
    \end{pmatrix},
\end{multline}
where
\begin{align}
    u & =  \sin^2\left(\frac{\sqrt{3}}{2}D\tau\right), \notag \\
    v & =  \cos^2\left(\frac{\sqrt{3}}{2}D\tau\right), \notag \\
    w & =  \sin(b)\sin\left(\sqrt{3}D\tau\right). 
\end{align}

An analogous calculation for the matrices $\rho^{1/2}_\mathrm{HT} (\tau)$ and $\rho^{1/2}_\mathrm{LT} (\tau)$ using Eqs.   (\ref{eq:liouvile_sol}), (\ref{eq:rho_LT_init}), and (\ref{eq:rho_HT_init}) yields

\begin{equation}
    \label{eq:rho_HT_eval_1_2}
    \rho_\mathrm{HT}^{1/2}(\tau) = 
    \begin{pmatrix}
            \frac 1 2  
        & 
            0  
        \\
            0
        & 
            -\frac 1 2   
    \end{pmatrix},
\end{equation}

\begin{equation}
\label{eq:rho_LT_eval_1_2}
    \rho_\mathrm{LT}^{1/2}(\tau) = \frac 1 Z 
    \begin{pmatrix}
            e^{\frac b 2}
        & 
            0  
        \\
            0
        & 
            e^{-\frac b 2}  
    \end{pmatrix}.
\end{equation}

Only the MQ NMR coherences of the zeroth and plus/minus second orders appear in the considered systems. These intensities can be calculated with Eqs.   (\ref{eq:coherence_k_s}) and (\ref{eq:rho_HT_eval_3_2}) - (\ref{eq:rho_LT_eval_1_2}):
\begin{align}
    \label{eq:analit_res_coherence}
    J_0(\tau) & = \dfrac{2\cos^2(\sqrt{3}D\tau) + 1}{3}, \notag \\
    J_{\pm 2}(\tau) & = \dfrac{\sin^2(\sqrt{3}D\tau)}{3} .
\end{align}

\begin{figure}
    \centering
    \includegraphics{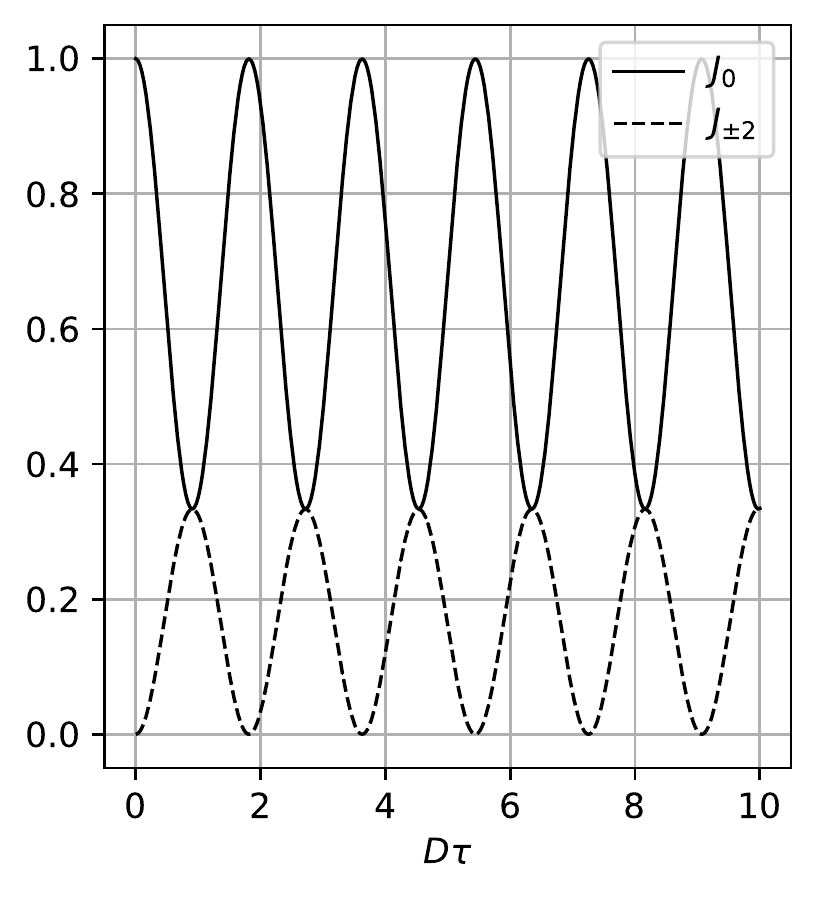}
    \caption{Intensities of MQ NMR coherences $J_n, \quad n=0, 2$ in a nanopore with $N = 3$.}
    \label{fig:exact_j}
\end{figure}

One can check that the sum of the intensities of Eq.   (\ref{eq:analit_res_coherence}) equals 1 independently of $\tau$ in accordance with Eq.   (\ref{eq:sum_of_coherence}). The profiles of the calculated intensities $J_n(\tau)$, n=0,2 are shown in Fig.\ref{fig:exact_j}.

\section{The second moment of the MQ NMR spectrum of the system of equivalent spins in the nanopore}
\label{sec:second_moment}

Generally speaking, the MQ NMR signal $G(\tau, \phi)$ of Eq. (\ref{eq:signal}) is not an out-of-time-ordered correlator (OTOC) \cite{otoc_to_enanglement_via_mqcoh} because it contains different matrices $\rho_\mathrm{LT}(\tau)$ and $\rho_\mathrm{HT}(\tau)$. The signal is an OTOC only in the high temperature approximation when $G(\tau, \phi)$ can be represented as

\begin{equation}
    \label{eq:otoc_HT}
    G(\tau, \phi) = \frac b Z\mathrm{Tr}\left\{ 
    e^{i \phi I_z} 
    \rho_\mathrm{HT} (\tau) 
    e^{-i \phi I_z} 
    \rho_\mathrm{HT}(\tau) 
    \right\}.
\end{equation}
At the same time, we can generalize the signal $G(\tau, \phi)$ of Eq. (\ref{eq:signal}) that it reduces to OTOC at arbitrary temperatures. For this end, one should average the signal after three periods of the MQ NMR experiment \cite{mq_nmr_experiment} over the initial low-temperature density matrix of Eq. (\ref{eq:rho_eq}).

Then 
\begin{multline}
    \label{eq:otoc_LT}
    G_\mathrm{LT}(\tau, \phi) = \\
  \mathrm{Tr}\left\{ 
        e^{i H_{\mathrm{MQ}} \tau} 
        e^{i \phi I_z} 
        e^{-i H_{\mathrm{MQ}} \tau} 
        \rho_{\mathrm{eq}}
        e^{i H_{\mathrm{MQ}}}
        e^{-i \phi I_z} 
        e^{-i H_{\mathrm{MQ}} \tau} 
        \rho_{\mathrm{eq}}
     \right\} = \\
   \mathrm{Tr}\left\{
         e^{i \phi I_z} 
         e^{-i H_\mathrm{MQ} \tau} 
         \rho_\mathrm{eq} 
         e^{i H_\mathrm{MQ} \tau} 
         e^{-i \phi I_z} 
         e^{-i H_\mathrm{MQ} \tau} 
         \rho_{\mathrm{eq}}
         e^{i H_\mathrm{MQ} \tau} 
     \right\} = \\
   \mathrm{Tr}\left\{ 
         e^{i \phi I_z} 
         \rho_\mathrm{LT} (\tau) 
         e^{-i \phi I_z} 
         \rho_\mathrm{LT} (\tau) 
     \right\}.
\end{multline}
It is evident from Eq. (\ref{eq:otoc_LT}) that $G_\mathrm{LT}(\tau, \phi)$ is OTOC at arbitrary temperatures.

The normalized intensities of the MQ NMR coherences for the correlator of Eq. (\ref{eq:otoc_LT}) can be written as 

\begin{equation}
    \label{eq:j_lt}
    J_{\mathrm{LT}, n}(\tau) = 
    \frac{\mathrm{Tr} \left\{ 
        \rho_{\mathrm{LT},n}(\tau)
        \rho_{\mathrm{LT}, -n}(\tau)
    \right\}}
    {\mathrm{Tr}\left\{\rho^2_\mathrm{eq}\right\}},             
\end{equation}
and a simple calculation yields

\begin{equation}
    \label{eq:j_lt_norm}
  \mathrm{Tr}\left\{\rho^2_\mathrm{eq}\right\} = 
    \frac{2^N \cosh^N(b)}{Z^2}.
\end{equation}
We call the coherences of Eq. ~(\ref{eq:j_lt}) the reduced multiple coherences.

In particular, the intensities of the reduced MQ NMR coherences for a system of $N=3$ spins are 
\begin{align}
    \label{eq:j_lt_3}
    J_{\mathrm{LT}, 0}(\tau) & = 1 - \frac 1 2 \tanh^2(b)\sin^2(\sqrt 3 D \tau), \notag \\
    J_{\mathrm{LT},\pm 2}(\tau) & = \frac 1 4 \tanh^2(b)\sin^2(\sqrt 3 D \tau)
\end{align}

The sum of the intensities of Eq. ~(\ref{eq:j_lt_3}) is again 1, as in the previous section~(\ref{sec:exact_sol}). However, the normalized intensities of Eq. ~(\ref{eq:j_lt_3})  depend now on the temperature, although the intensities of Eq. ~(\ref{eq:analit_res_coherence}) do not manifest such dependence. This is very important for the further analysis.

The second moment (dispersion) $M_2(\tau)$ of the distribution of the  reduced MQ NMR coherences $J_{LT, n} (\tau)$ can be expressed \cite{growrh_of_mqcoh} as

\begin{equation}
    \label{eq:dispersion}
    M_2(\tau) = \sum\limits_n n^2 J_{LT, n} (\tau).
\end{equation}

It was shown \cite{otoc_to_enanglement_via_mqcoh} that $2M_2(\tau)$ of Eq. (\ref{eq:dispersion}) determines a lower bound on the quantum Fisher information $F_{Q}$ \cite{qmetrology_for_qinfo,qmetrology_nonclassiscal_state} and $2M_2(\tau) \leq N^2$ \cite{fisher_and_entanglement}.
The numerical calculations presented in the following Section confirm this inequality.
\par
We give now a definition of many-particle entanglement  \cite{fisher_and_entanglement}. 
A pure state is $k$-particle entangled, if it can be written as a product \mbox{$\left| \Psi_\mathrm{k-ent} \right\rangle = \otimes^M_{l=1} \left| \Psi_l \right\rangle$}, where $\left| \Psi_l \right\rangle$ is a state of $N_l$ particles \mbox{($\sum\limits_{l=1}^M N_l = N$)}, each  $\left| \Psi_l \right\rangle$ does not factorize, and the maximal $N_l \geq k$. A generalisation for mixed states is straightforward \cite{fisher_and_entanglement}.
It was also ascertained \cite{qmetrology_for_qinfo,qmetrology_nonclassiscal_state} that, if

\begin{equation}
    \label{eq:fisher_criteria}
    F_{\mathrm{Q}} > mk^2 + (N-mk)^2,
\end{equation}
where $m$ is the integer part of $N/k$, then we have a $(k+1)$-particle entangled state in the system \cite{qmetrology_for_qinfo,qmetrology_nonclassiscal_state}.

Thus, we obtain a possibility to study the many-particle entanglement in a system of spin-carrying molecules (atoms) coupled with the DDIs in a nanopore.
The temperature dependence of the many-particle entanglement can also be investigated. 
At high temperatures, the intensities of the MQ NMR coherences can be investigated experimentally with usual MQ NMR experiments \cite{mq_nmr_experiment}. The results of the numerical analysis of the many-particle entanglement in the system of spin-carrying molecules (atoms) are presented in the following section.

\section{The temperature dependence of the many-particle entanglement}
\label{sec:entanglement}
\begin{figure}
    \centering
    \includegraphics{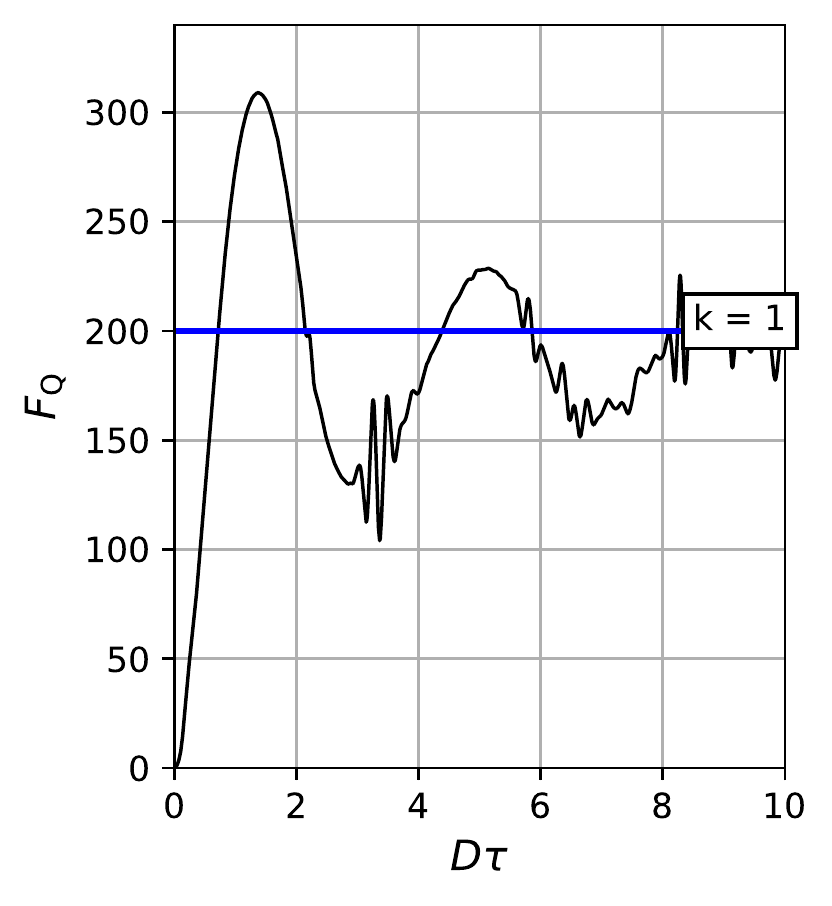}
    \caption{The dependence of the lower bound for the quantum Fisher information $F_Q = 2M_2(\tau)$ on the dimensionless time $D\tau$ at ${T=2.4\cdot10^{-1}}$~K $(b=0.1)$. 
    The inequality~(\ref{eq:fisher_criteria}) yields the region of pair entanglement $(k+1=2)$. The region is above the horizontal line. 
    }
    \label{fig:m2_t_b01}
\end{figure}

We use  the second moment $M_2$ of Eq. ~(\ref{eq:dispersion}) for the investigation of the many-particle entanglement in a spin system of 201 spins. 
The intensities of the reduced MQ NMR coherences are determined by Eqs.  ~(\ref{eq:j_lt},~\ref{eq:j_lt_norm}) both at high ($b < 1$) and low ($b > 1$) temperatures. 
We start from $b = 0.1$, which corresponds to the temperature ${T= 2.4\cdot 10^{-1}}$~K at the Larmor frequency $\omega_0 = 2\pi\cdot 500\cdot10^6$ s$^{-1}$ (Fig.~\ref{fig:m2_t_b01}). 
The inequality~(\ref{eq:fisher_criteria}) can be satisfied only when $k=1$ (the horizontal line on Fig.~\ref{fig:m2_t_b01}). 
This means pair entanglement is possible in the high-temperature case \cite{lab:mq_mnr_qinfo_2012}. 
\par
\begin{figure}
    \centering
    \includegraphics{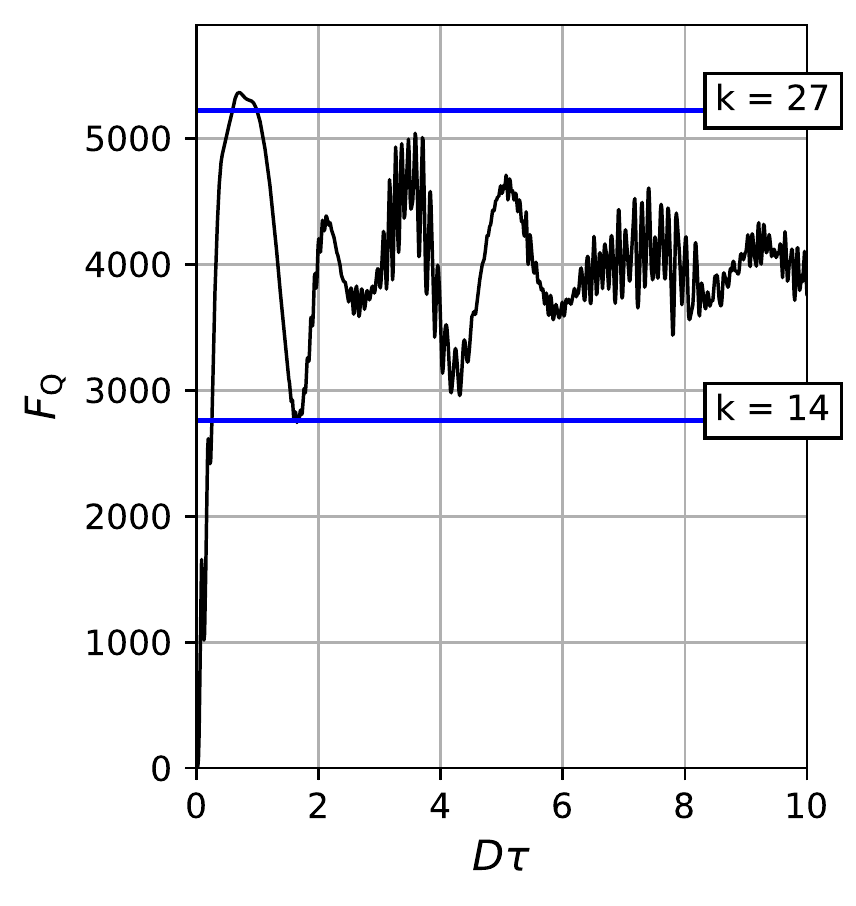}
    \caption{The dependence of the lower bound for the quantum Fisher information on the dimensionless time $D\tau$ at ${T=4.8\cdot10^{-2}}$~K $(b=0.5)$. The region of the many-spin entanglement is a strip bounded by the horizontal lines with $k=14$ and $k=27$.}
    \label{fig:m2_t_b05}
\end{figure}

\begin{figure}
    \centering
    \includegraphics{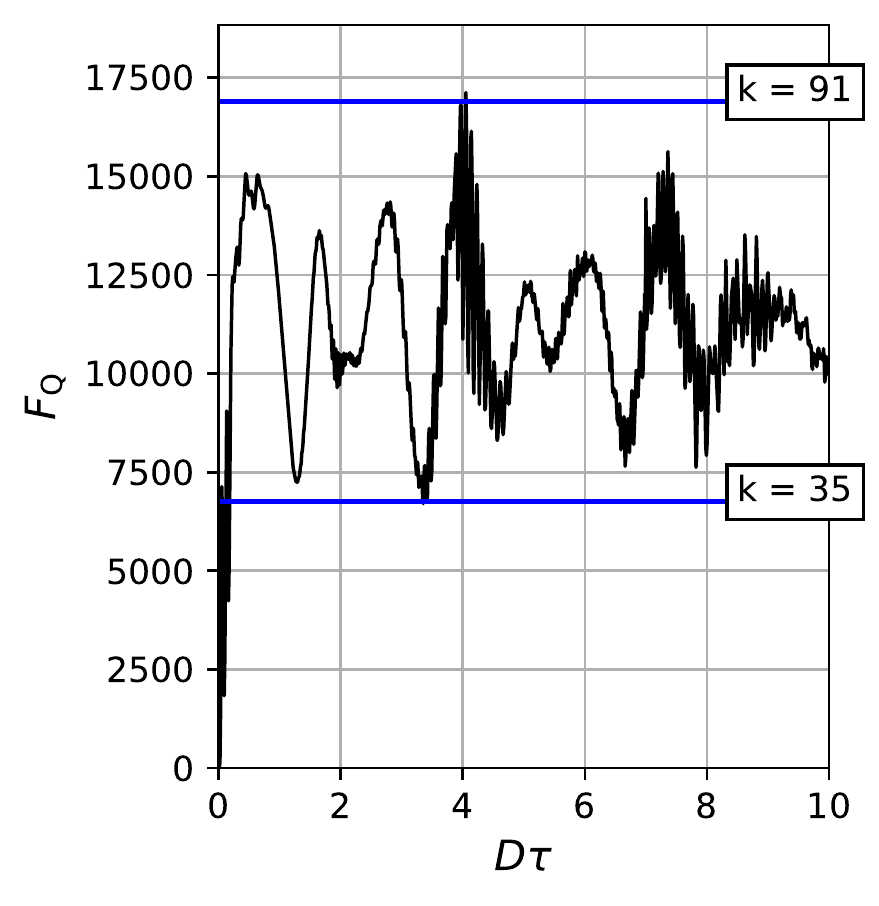}
    \caption{The dependence of the lower bound for the quantum Fisher information on the dimensionless time $D\tau$ at ${T=2.4\cdot10^{-2}}$~K $(b=1)$. The horizontal lines bound the strip with the many-spin entanglement.}
    \label{fig:m2_t_b1}
\end{figure}

 
At the temperature ${4.8\cdot10^{-2}}$~K $(b=0.5)$ one can see a strip (Fig.~\ref{fig:m2_t_b05}), in which the inequality~(\ref{eq:fisher_criteria}) can be satisfied when $14 \leq k \leq 27$.
Thus, there is many-spin entanglement in spin clusters consisting of 15 to 28 spins at the temperature  ${4.8\cdot10^{-2}}$~K. When the temperature decreases, the width of the strip, where many-spin entanglement exists, increases. At the temperature ${2.4\cdot10^{-2}}$~K $(b=1)$ (Fig.~\ref{fig:m2_t_b1}), in such a strip, the number of the entangled spins can range from 36 to 92.
\par
\begin{figure}
    \centering
    \includegraphics{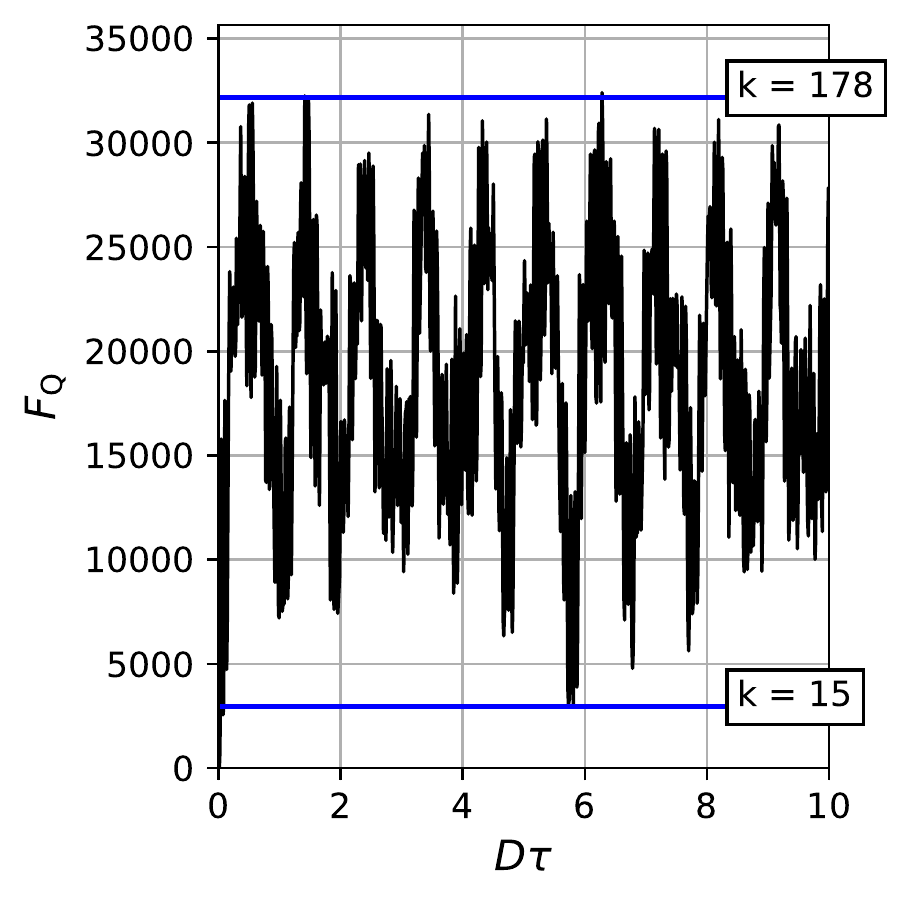}
    \caption{The dependence of the lower bound for the quantum Fisher information on the dimensionless time $D\tau$ at ${T=6.856\cdot10^{-3}}$~K $(b=3.5)$.
    Almost all spins (up to $179$ of $201$) can be a part of the entangled cluster.}
    \label{fig:m2_t_b3.5}
\end{figure}

Finally, at the temperature ${T= 6.856\cdot10^{-3}}$~K $(b=3.5)$ (Fig.~\ref{fig:m2_t_b3.5}), almost all spins (up to 179 of 201) are entangled. Entanglement exists during  the evolution process except a short initial period of time. 
\par
\begin{figure}
    \centering
    \includegraphics{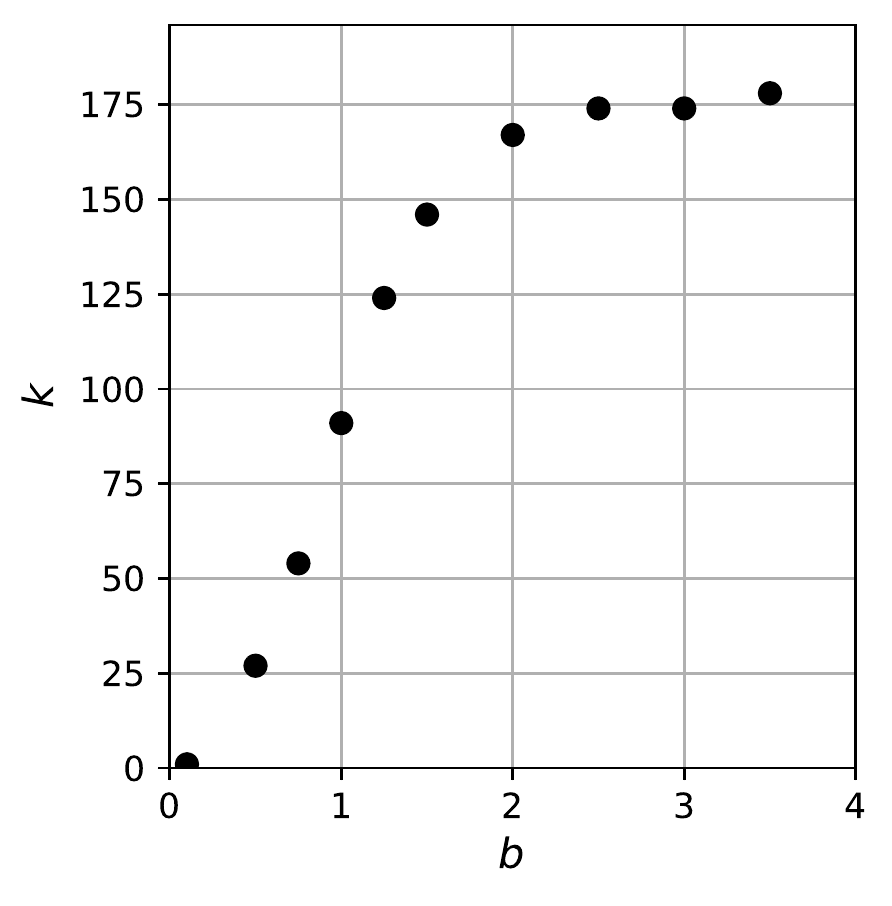}
    \caption{The dependence of the number of the entangled spins on the parameter $b = \frac{\hbar\omega_0}{kT} $.}
    \label{fig:k_b}
\end{figure}

 
Fig.\ref{fig:k_b} demonstrates that the number of the entangled spins increases when the temperature decreases.
\par
Thus, the suggested model of a nanocavity filled with spin-carrying atoms (molecules) allows us to investigate many-spin entanglement and its dependence on the temperature.

\section{Conclusion} 
\label{sec:conslusions}
We investigated many-particle entanglement in MQ NMR spectroscopy using a nanocavity filled with spin-carrying atoms (molecules). 
We developed a theory of MQ NMR in a nanocavity at low temperatures. 
The theory is based on the idea that  molecular diffusion is substantially faster than the time of the spin flip-flop processes. 
As a result, the problem is reduced to a system of equivalent spins [23, 25], which can be analyzed in the basis of the common eigenstates of the total spin angular momentum and its projection on the external magnetic field. 
Since there is a connection between the second moment (dispersion) of the distribution of the MQ NMR intensities and many-spin entanglement [17], we extracted information about many-spin entanglement from the MQ NMR spectrum. The temperature dependence of many-spin entanglement was also investigated.
\par
The main lesson consists in significant growth of many-particle entanglement at low temperatures. 
All or almost all spins are entangled at the dimensionless temperature $\frac{1}{b}$ of the order of 1. 
This suggests that $k$-entangled states with large $k$ emerge in a typical MQ NMR system at low temperatures. 
This is particularly interesting given the absence of entanglement in the initial state. We expect such behavior to be typical for MQ NMR. 
\par
We can conclude that MQ NMR spectroscopy is an effective method for the investigation of many-spin entanglement and the spreading of MQ correlations inside many-spin systems. It can be used for experimental investigations of quantum information processing in solids (note a related study of decoherence in liquids \cite{HOU2017863}).
\par

\begin{acknowledgments}
This work was performed as a part of the state task, state registration No.0089-2019-0002. This work is supported by the Russian Foundation for Basic Research, grant No.19-32-80004. I.L. acknowledges support from Foundation for the Advancement of Theoretical Physics and Mathematics “BASIS” No.19-1-5-130-1.  
\end{acknowledgments}

\bibliographystyle{apsrev4-1}
\bibliography{reference}

\end{document}